\documentclass[iop,twocolappendix]{emulateapj}

\usepackage{color}
\usepackage{natbib}
\usepackage{graphicx}
\usepackage{subfigure}
\usepackage{multirow}
\newcommand{\mini}{\mbox{$M_{\rm i}$}}

\newcommand{\Mhef}{\mbox{$M_{\rm HeF}$}}
\newcommand{\Mto}{\mbox{$M_{\rm TO}$}}
\newcommand{\diff}{\mbox{${\rm d}$}}
\newcommand{\sub}[1]{\mbox{$_{\rm #1}$}}

\newcommand{\Mbol}{\mbox{$M_{\rm bol}$}}

\newcommand{\feh}{\mbox{\rm [{\rm Fe}/{\rm H}]}}
\newcommand{\mh}{\mbox{\rm [{\rm M}/{\rm H}]}}
\newcommand{\Msun}{\mbox{$M_{\odot}$}}
\newcommand{\Lsun}{\mbox{$L_{\odot}$}}
\newcommand{\Mi}{\mbox{$M\sub{i}$}}
\newcommand{\Teff}{\mbox{$T_{\rm eff}$}}

\newcommand{\logl}{\mbox{$\log L/L_{\odot}$}}

\newcommand{\beq}{\begin{equation}}
\newcommand{\eeq}{\end{equation}}
\newcommand{\beqa}{\begin{eqnarray}}
\newcommand{\eeqa}{\end{eqnarray}}

\slugcomment{To appear in ApJ}

\shorttitle{Boosting the AGB of MC clusters}
\shortauthors{Girardi et al.}

\begin{document}


\title{The insidious boosting of TP-AGB stars in intermediate-age Magellanic Cloud clusters}

\author{
L\'eo Girardi\altaffilmark{1}, 
Paola Marigo\altaffilmark{2}, 
Alessandro Bressan\altaffilmark{3}, 
Philip Rosenfield\altaffilmark{4}
}
\altaffiltext{1}{Osservatorio Astronomico di Padova -- INAF, 
  Vicolo dell'Osservatorio 5, I-35122 Padova, Italy}
\altaffiltext{2}{Dipartimento di Fisica e Astronomia Galileo Galilei,
  Universit\`a di Padova, Vicolo dell'Osservatorio 3, I-35122 Padova, Italy}
\altaffiltext{3}{SISSA, via Bonomea 365, I-34136 Trieste, Italy}
\altaffiltext{4}{Department of Astronomy, University of Washington, Box 351580, Seattle, WA 98195, USA}

\begin{abstract}
In the recent controversy about the role of TP-AGB stars in evolutionary population synthesis (EPS) models of galaxies, one particular aspect is puzzling: TP-AGB models aimed at reproducing the lifetimes and integrated fluxes of the TP-AGB phase in Magellanic Cloud (MC) clusters, when incorporated into EPS models, are found to overestimate, to various extents, the TP-AGB contribution in resolved star counts and integrated spectra of galaxies. In this paper, we call attention to a particular evolutionary aspect, linked to the physics of stellar interiors, that in all probability is the main cause of this conundrum.
As soon as stellar populations intercept the ages at which RGB stars first appear, a sudden and abrupt change in the lifetime of the core He-burning phase causes a temporary ``boost'' in the production rate of subsequent evolutionary phases, including the TP-AGB. For a timespan of about 0.1~Gyr, triple TP-AGB branches develop at slightly different initial masses, causing their frequency and contribution to the integrated luminosity of the stellar population to increase by a factor of $\sim\!2$. The boost occurs for turn-off masses of $\sim\!1.75\Msun$, just in the proximity of the expected peak in the TP-AGB lifetimes (for MC metallicities), and for ages of $\sim\!1.6$~Gyr. Coincidently, this relatively narrow age interval happens to contain the few very massive MC clusters that host most of the TP-AGB stars used to constrain stellar evolution and EPS models. This concomitance makes the AGB-boosting particularly insidious in the context of present EPS models. As we discuss in this paper, the identification of this evolutionary effect brings about three main consequences.
First, we claim that present estimates of the TP-AGB contribution to the integrated light of galaxies derived from MC clusters, are biased towards too large values.
Second, the relative TP-AGB contribution of single-burst populations falling in this critical age range cannot be accurately derived by approximations such as the fuel consumption theorem, that ignore, by construction, the above evolutionary effect.
Third, a careful revision of AGB star populations in intermediate-age MC clusters is urgently demanded, promisingly with the aid of detailed sets of stellar isochrones.
\end{abstract}




\section{Introduction}
\label{intro}

It is well established that a sizable fraction of the integrated light of stellar populations comes from the thermally pulsing asymptotic giant branch (TP-AGB) phase \citep[][]{Frogel_etal90}. However, the size of this fraction has been subject of much discussion in the recent literature, with evolutionary population synthesis (EPS) models of galaxies favouring either ``heavy'' \citep[][]{Maraston05, Maraston_etal06} or ``light'' \citep{Kriek_etal10, Zibetti_etal13} flux contributions from TP-AGB stars. Efforts to account for the TP-AGB contribution to the integrated light of galaxies are based on two different techniques:

1) EPS models based on the isochrone method \citep{CharlotBruzual91, BruzualCharlot93} start by adopting the best-available sets of evolutionary tracks and isochrones including the TP-AGB phase. Since this phase is notoriously challenging to model -- due to difficulties and uncertainties in the description of mixing, energy transport by convection, mass loss, and numerical aspects -- at present there is no set of widely-accepted TP-AGB model grids in the literature. Therefore, in general, the choice falls on models which at least try to reproduce basic observables of TP-AGB stars in the Magellanic Clouds \citep[MCs; e.g.][]{MarigoGirardi07, Marigo_etal08, WeissFerguson09}. This approach has been adopted in the popular \citet{BruzualCharlot03} and \citet{Conroy_etal09} models, with some subtle technical differences and a posteriori corrections -- to consider e.g.\ circumstellar dust \citep{GonzalezLopezlira_etal10} or shifts in the $\Teff$ and $L$ of TP-AGB models \citep{ConroyGunn10}.

2) EPS models by \citet{Maraston98, Maraston05}, based on the fuel consumption theorem \citep[cf.][hereinafter FCT]{RenziniBuzzoni86}, bypass the use of TP-AGB evolutionary tracks and stellar isochrones, since they are rooted in the belief that detailed modelling of the TP-AGB evolution is hopelessly uncertain to be useful in this context. Each single-burst stellar population is simply assigned, as a function of age, the TP-AGB ``fuel'' that appears to best reproduce the TP-AGB luminosity contribution measured in MC clusters. Besides the integrated fuel as a function of stellar age, the FCT method employs other prescriptions, such as the fraction of fuel burnt by C- and M-type stars, and some rough dependence on metallicity.

\begin{figure*}
\includegraphics[width=\columnwidth]{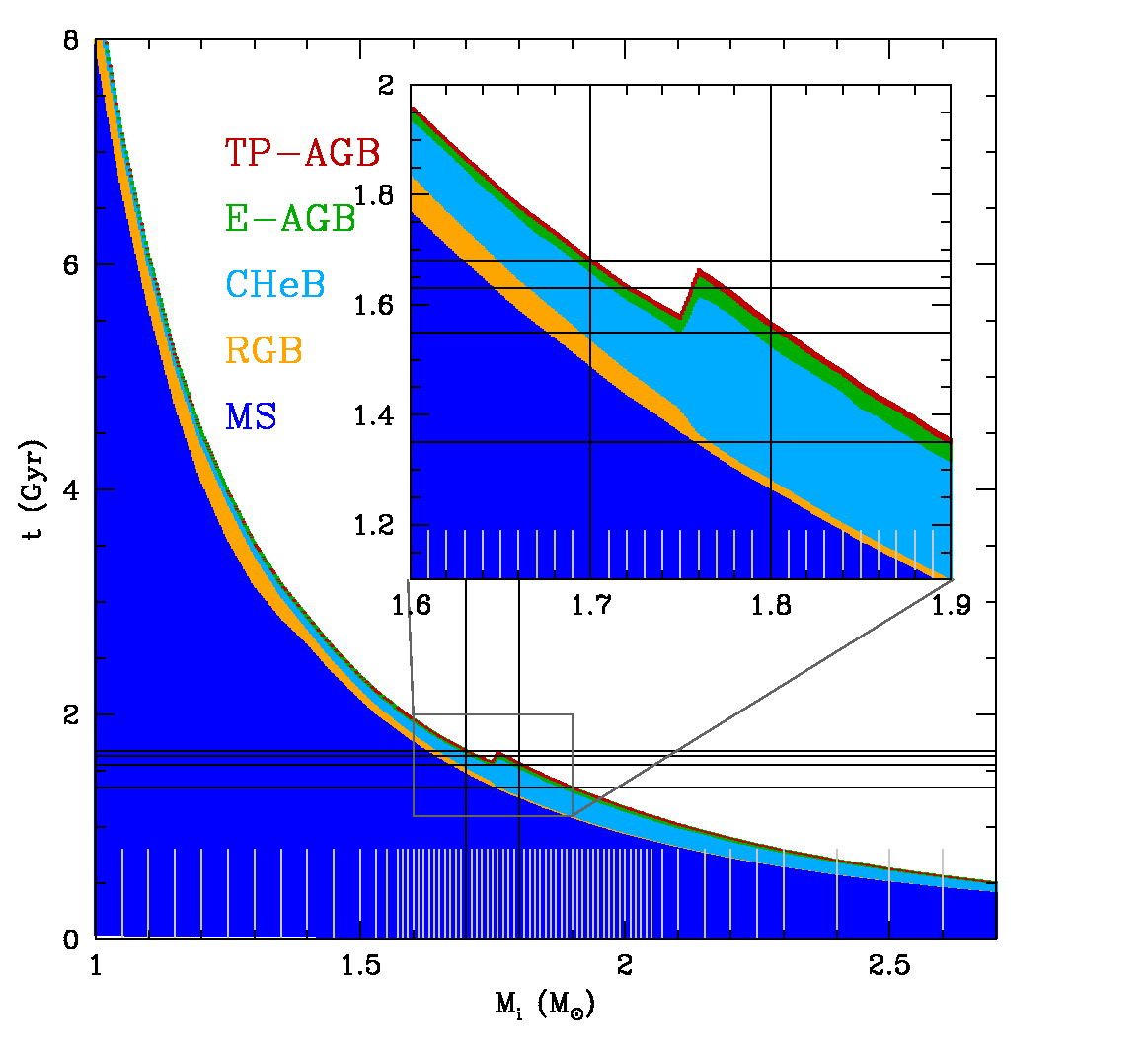}
\includegraphics[width=\columnwidth]{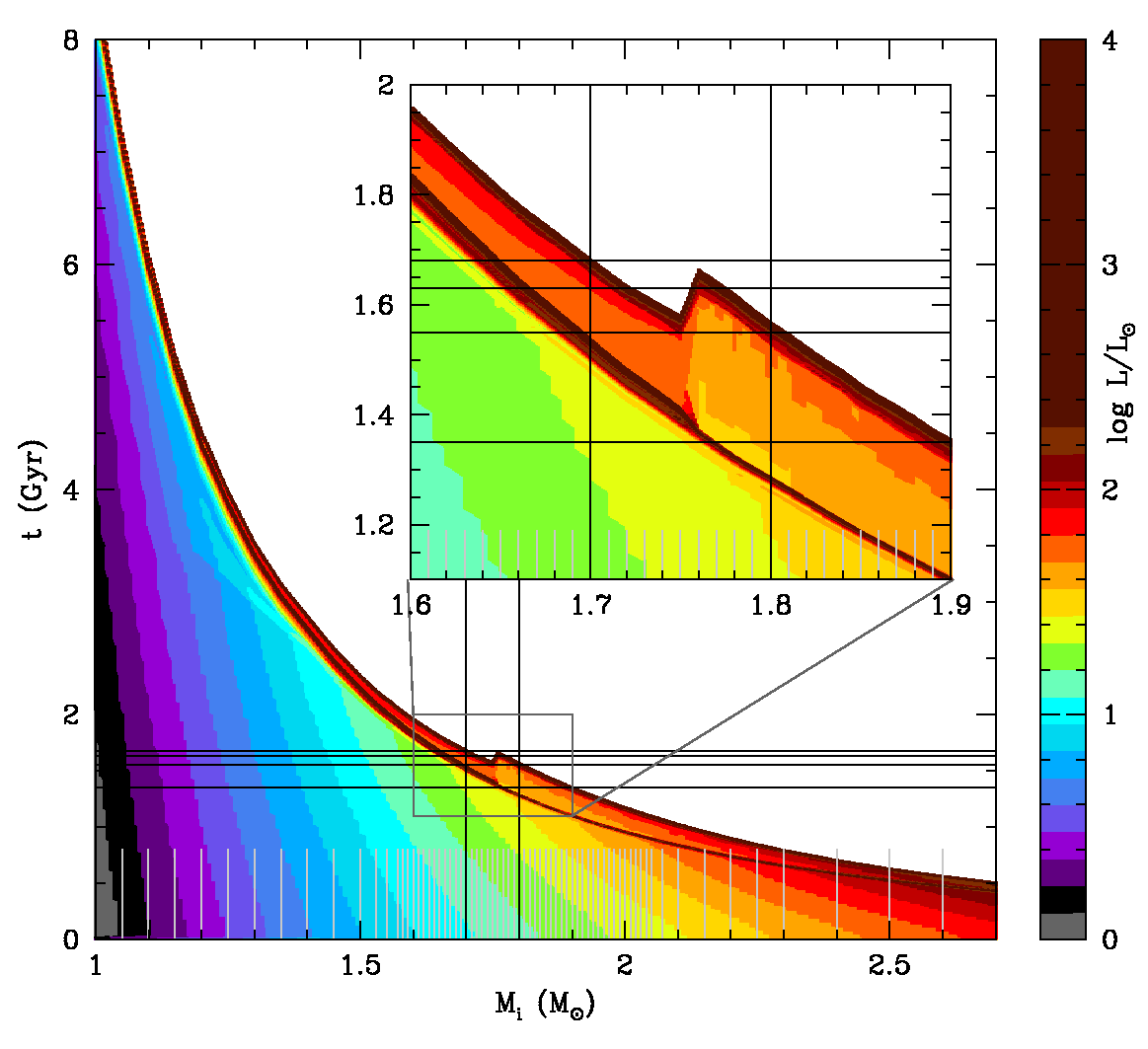}
\caption{The \mini\ vs. $t$ plot, in which vertical lines represent stellar evolutionary tracks. Small vertical grey bars indicate the initial masses of tracks which were actually computed, from which all others are derived via interpolation. The inset zooms into the crucial region that is discussed in this paper. In the \textbf{left panel}, different sections of the tracks are colored according to their evolutionary stage. As expected, single evolutionary tracks intercept each evolutionary phase only once. One can notice the marked changes in the lifetimes of evolutionary phases subsequent to the MS, at $\Mi\simeq1.75$~\Msun. On the \textbf{right panel}, the same tracks are coloured according to their \logl. It is evident how the luminosity of the red clump and subsequent evolutionary phases changes abruptly in the proximity of \Mhef. The dark horizontal lines mark the position of several isochrones discussed in this paper.}
\label{fig_taum}
\end{figure*}

It is somewhat puzzling and disturbing that, despite the efforts of calibrating these EPS models on TP-AGB data in the MC clusters (albeit with different approaches), when taken without their ``corrections'',  the same EPS models tend to overestimate the contribution of the TP-AGB to the spectral energy distributions of galaxies. 
For instance, \citet{Maraston05} models, based on the FCT recipe, show an excess of TP-AGB flux which is not observed in post-starburst galaxies \citep{Kriek_etal10, Zibetti_etal13}. Likewise, \citet{MarigoGirardi07} evolutionary tracks predict $\sim\!40\,\%$ more AGB stars than observed in a sample of nearby galaxies observed with HST, which translates to a factor $\sim\!2$ excess in their integrated near-infrared flux \citep{Melbourne_etal12, Johnson_etal13}.

In this study we analyse a specific evolutionary effect which we identify as the likely reason for this unexpected incongruity. We start describing in detail how the TP-AGB integrated luminosity is predicted to vary with age (Sect.~\ref{models}). We find a significant boosting of the TP-AGB contribution -- not predicted by the FCT -- at ages $t\sim\!1.6$~Gyr, that coincides with the age interval populated by some of the most massive clusters in the MCs (Sect.~\ref{controversy}). The far-reaching consequences and implications in the context of the TP-AGB calibration are discussed in Sect.~\ref{conclu}. 

\section{Stellar evolution models}
\label{models}

To explore the detailed contribution of TP-AGB stars to the integrated light of a SSP, we have computed a dense grid of stellar evolutionary tracks, assuming  an initial composition with metallicity $Z=0.006$, helium abundance $Y=0.259$, and a scaled-solar distribution of metals \citep[cf.][]{Caffau_etal11}. This corresponds to the case $\feh=\mh=-0.40$, which is suitable to describe intermediate-age clusters in the LMC.

Stellar evolutionary tracks were computed with the PARSEC code \citep{Bressan_etal12} until the first thermal pulse on the TP-AGB, and then followed with the COLIBRI code \citep{Marigo_etal13} until the complete ejection of the stellar envelope. Tracks are closely distributed in initial mass $\Mi$, with a spacing $\Delta\Mi=0.01$~\Msun\ in the proximity of the limiting maximum mass, $\Mhef$, for a star to develop an electron-degenerate He-core after the main sequence (MS; with $\Mhef=1.75$~\Msun\ in this case). For low-mass stars (those with $\Mi<\Mhef$), the evolution from the He-flash to the initial stage of quiescent He core-burning (CHeB) is skipped by means of a standard and well-tested algorithm that preserves the stellar core mass and chemical profile, while converting part ot the helium in the core into carbon to take into account the nuclear energy necessary to lift the electron degeneracy. Convective core overshooting is adopted with an efficiency as detailed in \citet{Bressan_etal12}. In addition, we completely suppress breathing pulses towards the end of CHeB phase, so as to avoid erratic track-to-track fluctuations in the CHeB lifetimes.

\subsection{The initial mass vs.\ age relation}

Evolutionary tracks are shown in the initial mass versus age (\Mi\ vs. $t$) plots of Fig.~\ref{fig_taum}, color-coded as a function of the evolutionary phase (left panel) or luminosity (right panel). The inverse power-law relation between the initial mass and MS lifetime is clearly recognizable in the plot, as is the presence of a long-lived red giant branch (RGB) for masses $\Mi\!<\!\Mhef$. Also evident is that the CHeB lifetime suddenly gets longer at $\Mi\!>\!\Mhef$, passing from the $\approx\!10^8$~yr typical of low-mass stars to at least twice this value at $\Mi\!\simeq\!1.8$~\Msun. This abrupt growth reflects the onset of CHeB at significantly lower core masses and luminosities, compared to the case of low-mass stars. These features are well-established and their consequences to the CHeB phase are thoroughly discussed in \citet{Girardi99}. Moreover, we notice how brief the typical lifetime of the TP-AGB phase (a few $10^6$~yr) is compared to the CHeB lifetime, and its relatively narrow dynamical range over the relevant mass interval. 

In Fig.~\ref{fig_taum} a few horizontal lines are drawn to highlight the evolutionary phases intersected by the stellar \textit{isochrones} at some selected ages:
\begin{itemize}
\item The youngest isochrone, of $t=1.35$~Gyr, crosses the MS, a very short section of ``RGB'' (actually, this is just the quick core contraction phase that precedes He-ignition in a non-degenerate core), a long section of CHeB, the early-AGB (E-AGB), and the TP-AGB. 
\item Similarly, the same phases are crossed by the 1.68-Gyr isochrone, with the difference that its RGB is somewhat longer (in terms of the spanned \Mi\ interval) and more luminous, while the CHeB is shorter.
\item At 1.55 Gyr, the isochrone intercepts both a section of low-mass and bright CHeB, and a section of higher-mass (and fainter) CHeB. This is the age range in which the most rapid changes in CHeB morphology take place.
\item At 1.63 Gyr, the isochrone crosses three distinct TP-AGB phases (as well as three E-AGBs), the first at $\sim\!1.72$~\Msun, the second (somewhat shorter) at $\sim\!1.76$~\Msun, and the third at  $\sim\!1.78$~\Msun. The second TP-AGB section raises quickly from below to above the 1.63~Gyr line, so it is likely less populated than the other two.
\end{itemize}
 
It is somewhat obvious from Fig.~\ref{fig_taum} that the numbers of TP-AGB stars will be ``boosted'' in isochrones with ages between 1.57 and 1.66~Gyr, since they will all contain at least two well populated TP-AGBs. Since the two branches have similar (and high) luminosities, their total integrated fluxes will be boosted as well. A rough estimate of the boosting factor comes from the integral of $\Mi$, weighted by the IMF, along the sections of the isochrone lines that correspond to the TP-AGBs. Given the limited range of masses involved, we can naively expect a boosting factor of $\sim\!2$ to be typical across this 0.1 Gyr-wide age interval.

\begin{figure}
\includegraphics[width=\columnwidth]{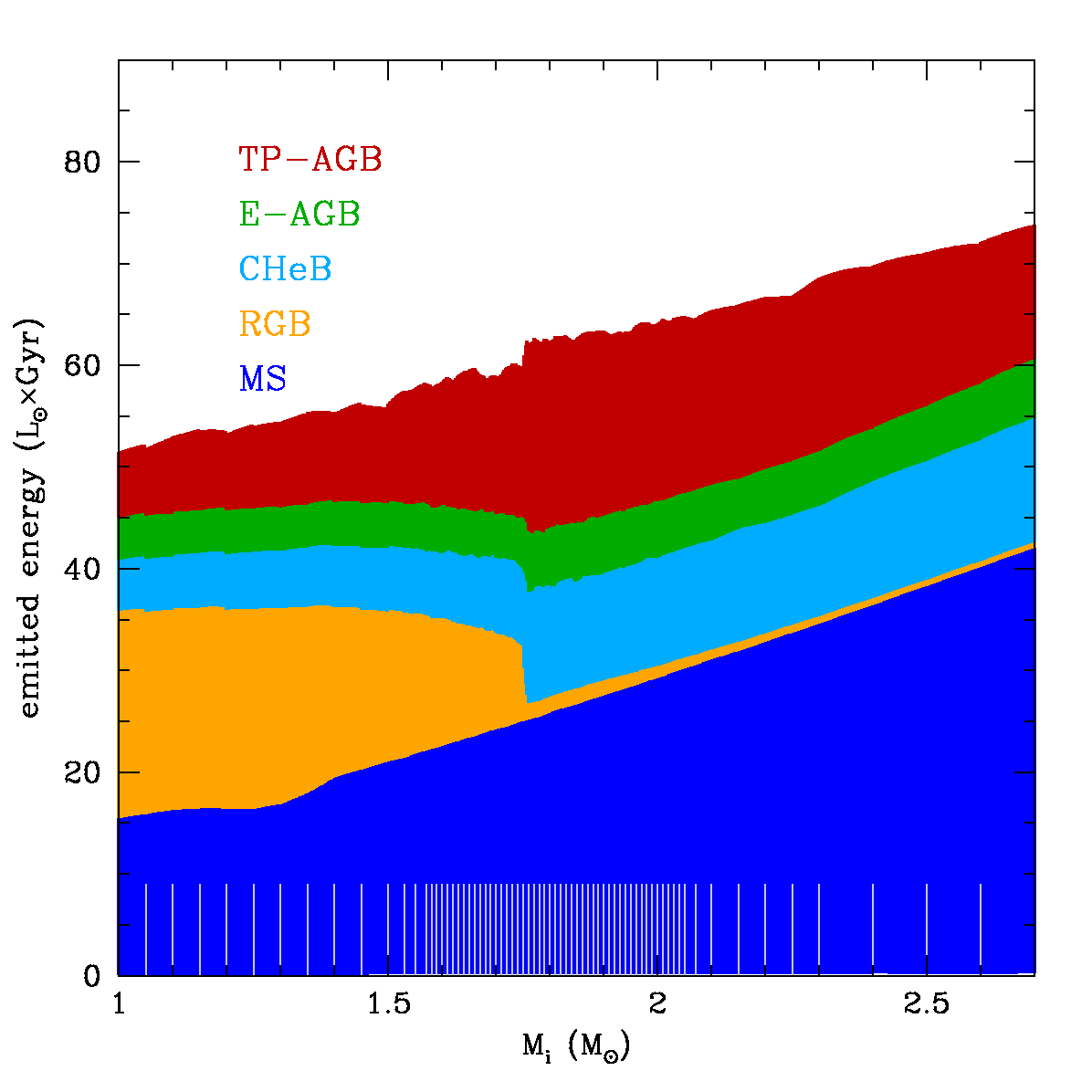}
\caption{The integrated emitted energy during the stellar life (from Eq.~\ref{eq_intfuel}), as a function of \Mi. 
Like in the previous Fig.~\ref{fig_taum}, different evolutionary stages are marked with different colors, and the vertical bars at the bottom mark the \Mi\ of the tracks which were actually calculated, rather than derived from interpolation.}
\label{fig_fuelm}
\end{figure}

\subsection{Integrated luminosities of TP-AGB stars: isochrones vs.\ fuel-consumption theorem}

Let us now look at this problem in terms of the luminosity contribution of TP-AGB stars, $L_{\rm TP-AGB}(t)$, to the integrated bolometric light of a single-burst stellar population (SSP) as a function of its age $t$. 

As a first step, let us consider the integrated emitted energy of stellar tracks from their birth line on the pre-MS, up to any given evolutionary stage $s$:
\begin{equation}
E^{\rm tot}_{\rm s}(\Mi)= \left. \int_{0}^{t_{\rm s}} L(t) \diff t \right|_{\Mi={\rm constant}}   \,\,\,\,.
\label{eq_intfuel}
\end{equation}
Figure~\ref{fig_fuelm} shows this quantity as a function of the initial stellar mass, as obtained from our set of evolutionary tracks, and it is color-coded as a function of the main evolutionary stages. We can appreciate that the total emitted energy is a quite well-behaved function of \Mi. The most notable feature in this plot is the presence of a significant contribution from the RGB at masses $\Mi\!<\!\Mhef$, which is counterbalanced by an increase in the CHeB and TP-AGB contributions at $\Mi\!>\!\Mhef$. Except for these aspects, the run of the emitted energy with mass is quite smooth and covers a remarkably limited range. For this particular set of tracks, the total post-MS emitted energy amounts to $\sim\!35-40$~\Lsun\,Gyr, with the TP-AGB contribution ranging from $\sim\!5$ to $\sim\!20$~\Lsun\,Gyr, depending on the stellar mass. 

\begin{figure*}
\includegraphics[width=\columnwidth]{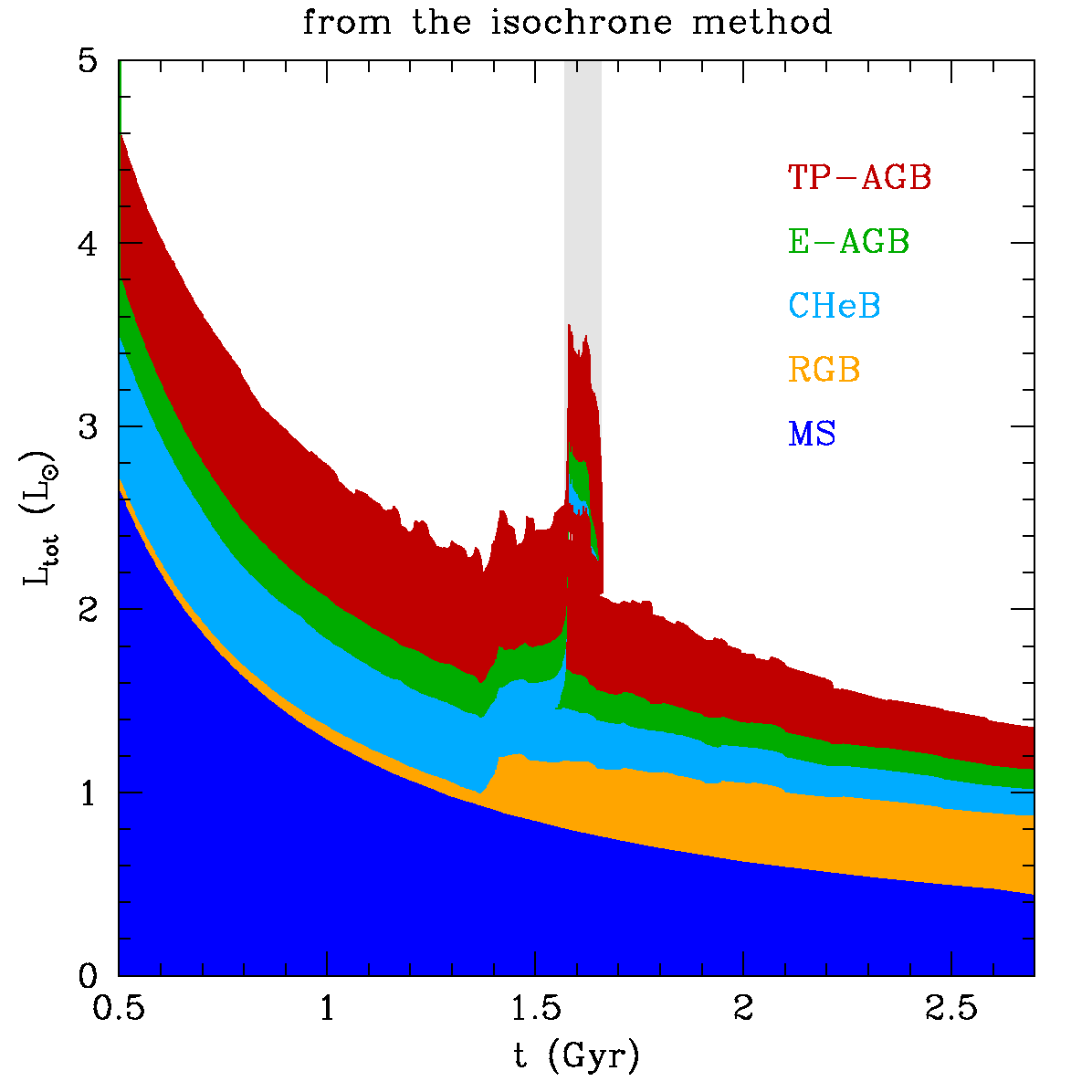}
\includegraphics[width=\columnwidth]{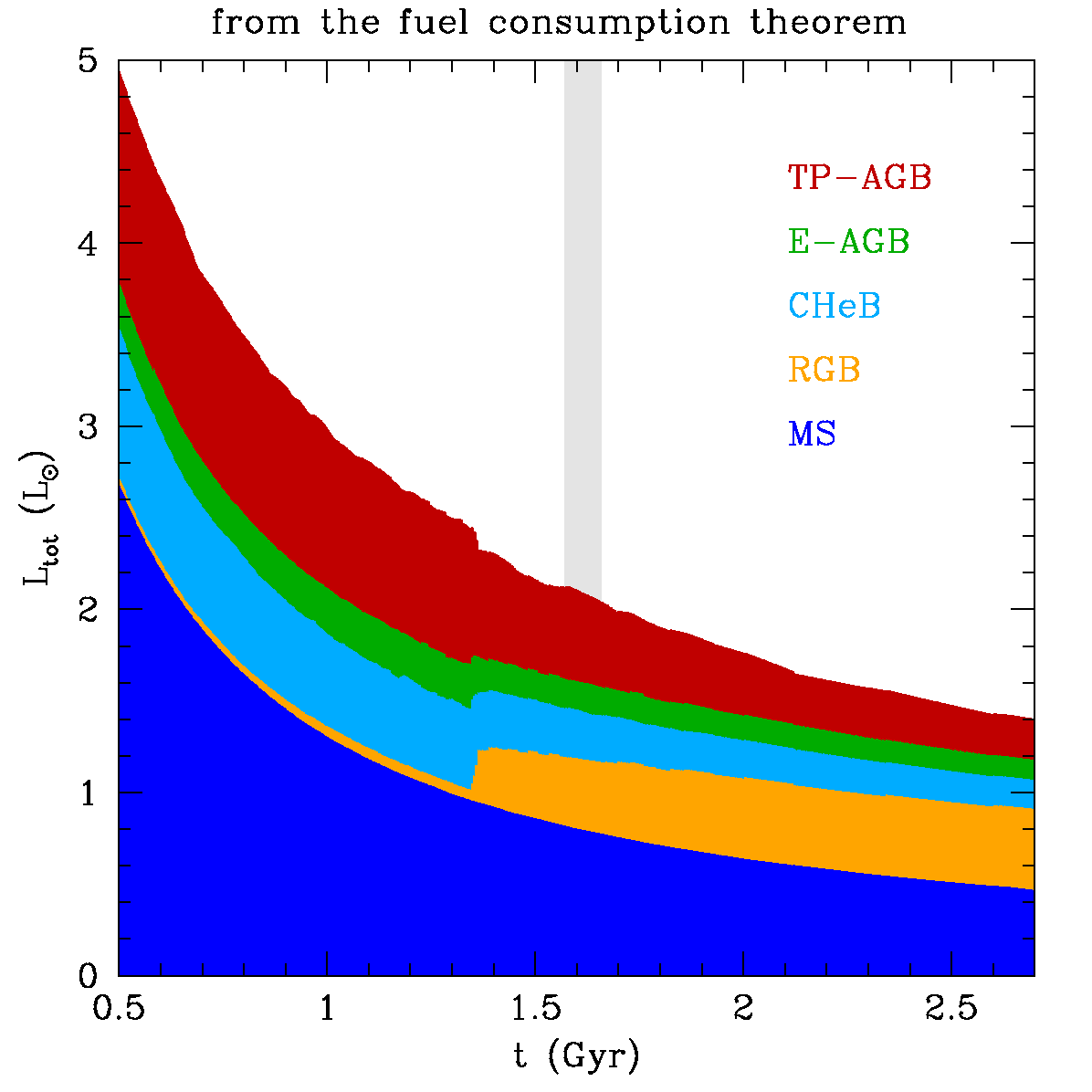}
\caption{ {\bf Left panel:} The evolution of integrated bolometric luminosity of a SSP as derived from our detailed isochrones (Eq.~\ref{eq_intfuelisoc}), as a function of age. Notice the presence of the AGB-boosting period at ages $\sim\!1.6$~Gyr, marked by the shaded grey area.  As in previous figures, different colors indicate the fraction of luminosity coming from different evolutionary stages. The {\bf right panel} instead shows the evolution of the luminosity as predicted by the FCT (Eq.~\ref{ltpagb_fct}), when applied to the same tracks that generated the isochrones in the left panel. The TP-AGB boosting period is absent in this case.}
\label{fig_fuelt}
\end{figure*}

Let us now look at the SSP integrated luminosity as derived from the isochrones,
\begin{equation}
L^{\rm tot}_{\rm s}(t)= \left. \int_{0}^{\Mi_{\rm s}} L(\Mi)\phi(\Mi) \diff \Mi \right|_{t={\rm constant}}   \,\,\,\,,
\label{eq_intfuelisoc}
\end{equation}
where $\phi(\Mi)$ initial mass function \citep[in this case the][with a low-mass cut at $\Mi\!>\!0.5$~\Msun]{Salpeter55}. This quantity is shown in the left panel of Fig.~\ref{fig_fuelt}, as a function of age, and again separating the contribution from different evolutionary stages. Notice the significant increase of the total TP-AGB luminosity contribution at ages between $1.57$ and 1.66~Gyr, the unequivocal evidence of the ``AGB boosting'' that originates from the multiple TP-AGBs at those ages. Moreover, notice that there is a period between 1.40~Gyr and 1.55~Gyr during which the increase in luminosity due to the appearance of the RGB is not compensated by a decrease in the TP-AGB luminosity, so that also in this interval there is a temporary but more modest increase in the total emitted luminosity. Only for ages older than 1.66~Gyr the evolution of the total luminosity seems to recover again the smooth trend defined by ages younger than 1.40~Gyr. 

The figure also contains some mild irregularities, especially at the end of the TP-AGB phase, which derive either from the small variations in the rate at which AGB stars are produced after the CHeB phase, or from the variable numbers of thermal pulses occurring in the single TP-AGB evolutionary tracks. Anyway, these fluctuations are small and not of concern here. The most significant point, in the context of this study, is the presence of the AGB-boosting period between $1.57$ and 1.66~Gyr.

As already mentioned, the AGB-boosting period derives from the sudden increase of the CHeB lifetime at $\Mi>\Mhef$, that causes single isochrones to cross multiple sections of the TP-AGB phase experienced by stars with slightly different initial masses. Approximations which do not take into account these different CHeB lifetimes and how they reflect into isochrones, are not expected to show this feature. In particular, the FCT \citep[cf.][]{RenziniBuzzoni86, Maraston05} approximates the evolution of $L^{\rm tot}_{\rm s}(t)$ with the following equation:
\begin{equation}
L^{\rm tot, FCT}_{\rm s}(t) = L^{\rm tot}_{\rm MS}(t) + \phi(M_{\rm TO})\, |\dot{M}_{\rm TO}| 
\int_{t_{\rm TO}}^{t_{\rm s}}  L_{M_{\rm TO}}(t) \diff t  \,\,\,\,,
\label{ltpagb_fct}
\end{equation}
where $L^{\rm tot}_{\rm MS}(t)$ is the integrated luminosity of the MS. All the subsequent evolutionary stages are described by a single track of mass equal to the turn-off one, $\Mto$, uniformly weighted by the evolutionary rate at which stars with $\Mi=\Mto$ leave the MS, 
$\phi(M_{\rm TO})\,|\dot{M}_{\rm TO}|$. 
The right panel of Fig.~\ref{fig_fuelt} shows the result of applying equation~\ref{ltpagb_fct}, where the functions $L_{M_{\rm TO}}(t)$ (equivalent to ``tables of fuel'') are obtained from the evolutionary tracks, and are shown in our previous Fig.~\ref{fig_fuelm}. We adopt the same layout as in the left panel, showing the contribution of every evolutionary phase to the integrated light with a different color. Evidently there is no sign of an AGB-boosting period in this case.

The reason of such a remarkable difference in the predictions between the isochrone and FCT methods can be easily caught considering that, to describe the post-MS phases of a SSP, the FCT assumes a perfectly one-to-one correspondence between age-index and mass-index (MS lifetime of a star with mass $M_{\rm TO}$), while the isochrone method populates each phase by a finite range of initial stellar masses.
It follows that the FCT approximation constrains any SSP to cross a given post-MS phase not more than once (exemplified by the vertical lines in Fig.~\ref{fig_taum}), while multiple crossings may, in principle, take place with the isochrone method (represented by e.g.\ the horizontal lines in Fig.~\ref{fig_taum}).
 
Of the two approaches shown in Fig.~\ref{fig_fuelt}, the one based on isochrones \citep[see also][]{CharlotBruzual91} is certainly best suited to represent the time evolution of single-burst stellar populations -- defined as generations of stars born at the same time and sharing the same age $t$ -- alike small-mass star clusters. It is also the one that turns out to capture most details of the SSP luminosity evolution.


\begin{figure*}
\includegraphics[width=0.9\textwidth]{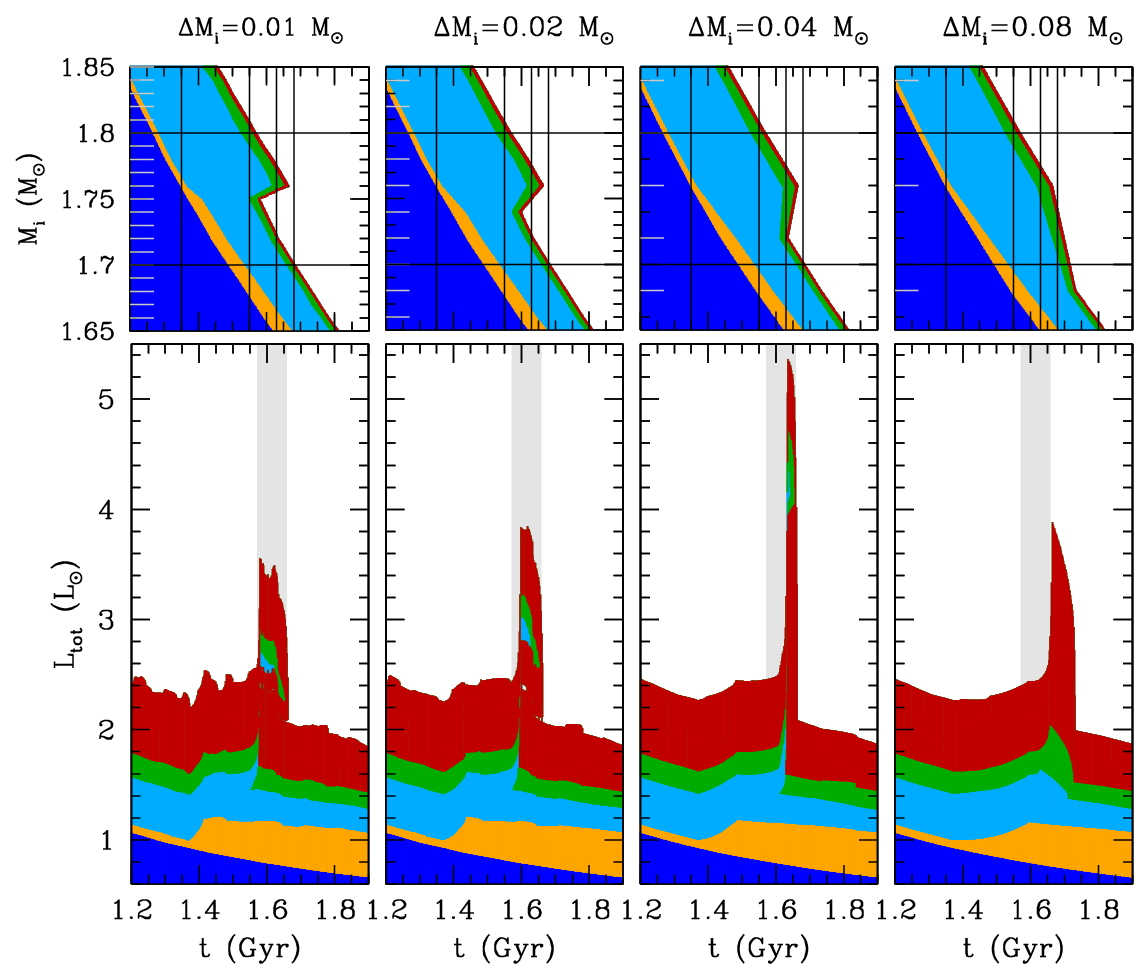}
\caption{The effect of reducing the mass resolution of the grid of evolutionary tracks, on the appearance of the AGB-boosting period. The panels from left to right show reduced versions of the left panels of Fig.~\ref{fig_taum}'s \mini\ vs.\ $t$ (top panels) and Fig.~\ref{fig_fuelt}'s integrated luminosity vs.\ $t$ plots(bottom panels). In both cases we detail the age interval from 1.2 to 1.9 Gyr, which completely covers the development of the RGB and AGB-boosting periods. The top panels are rotated w.r.t.\ those of Fig.~\ref{fig_taum}, so as to allow a direct comparison with the age scale in the bottom panels. Each couple of panels correspond to a given mass resolution $\Delta\Mi$, as indicated at the top line. The sequence from left to right shows that reducing the resolution of the grid of tracks causes changes in the amplitude and age limits of the AGB-boosting period, but simply does not eliminate it, neither reduce its total impact the evolution of the integrated luminosity.}
\label{fig_resolu}
\end{figure*}

\subsection{The role of the mass resolution}

This is not the first time that the AGB boosting at ages close to 1.6~Gyr has been noticed. Using isochrones derived from evolutionary tracks computed with a coarser mass resolution, \citet{GirardiBertelli98} already noticed a temporary increase in the TP-AGB production rate due to the flattening of the $t\sub{eHe}$ versus \Mi\ relation (where $t\sub{eHe}$ is the age at the end of the CHeB phase) close to \Mhef. The increase was in proportion to $\phi(M_{\rm eHe})\,|\dot{M}_{\rm eHe}|$, and appeared as a sharp, short-lived peak in the time evolution of integrated colors such as $V\!-\!K$. This feature has been present in many sets of the Padova isochrones distributed since 2000 \citep{Girardi_etal00, Marigo_etal08}\footnote{\url{http://stev.oapd.inaf.it/cmd}}, and often is considered to be a bug by external users. By conveniently sampling a grid of isochrone ages one could completely eliminate it from the data. The feature we identify and discuss in this work is essentially the same one, but this time it has been derived from a much more detailed and denser grid of stellar evolutionary tracks. 
What was seen as a sharp peak in the production of AGB stars appearing at a precise age by \citet{GirardiBertelli98}, it now appears as a triple TP-AGB over a relatively wide age interval (Fig.~\ref{fig_fuelt}). 

Let us give a closer look at this point. Fig.~\ref{fig_resolu} illustrates essentially the same features as in the previous Figs.~\ref{fig_taum} and \ref{fig_fuelt}, but now using grids of evolutionary tracks of progressively worse mass resolution. These degraded grids are obtained from the high-resolution grid already described, by progressively eliminating tracks in the vicinity of the 1.76~\Msun\ one. Interpolated tracks and isochrones are then obtained using exactly the same algorithms as before. Four different resolutions are presented, namely $\Delta\Mi=0.01$, 0.02, 0.04 and 0.08~\Msun, the first one being the original case already presented in previous Figs.~\ref{fig_taum} to \ref{fig_fuelt}. The presence of the 1.76~\Msun\ track ensures that, in all cases, we sample the maximum of He-burning lifetimes that occurs slightly above \Mhef.

For the $\Delta\Mi=0.02$ and $\Delta\Mi=0.04$~\Msun\ cases, the mass resolution is still good enough for a triple TP-AGB to appear in the isochrones (top-middle panels), although for a reduced age interval. As a consequence, the AGB-boosting period continues to appear in the integrated light (bottom-middle panels), but confined to a smaller age range, {\em and with a larger amplitude}. The larger amplitude is simply caused by the fact that TP-AGB stars from a wider range of initial masses now appear in a narrower range of isochrone ages. In the case of the $\Delta\Mi=0.02$~\Msun\ resolution, the effect is not dramatic, and the evolution of integrated light resembles very much the one obtained for the $\Delta\Mi=0.01$~\Msun\ case. For $\Delta\Mi=0.04$~\Msun, however, the AGB-boosting appears over an age range of just $\sim\!0.04$~Gyr, and with a amplitude about 2.5~times larger than the one seen at $\Delta\Mi=0.01$~\Msun.

For the $\Delta\Mi=0.08$~\Msun\ case, instead, the situation is apparently very different. There is no longer any single isochrone crossing three different TP-AGB sections (as seen in the top-right panel). But anyway, an intense AGB-boosting period still appears, and at slightly later ages than before. This happens because there is an age interval (between 1.66 and 1.73~Gyr) in which the isochrones cross a larger interval of \Mi\ while on the TP-AGB phase.  This is actually the situation which was found by \citet{GirardiBertelli98}, and explained by means of the higher production rate of AGB stars (proportional to  $\phi(M_{\rm eHe})\,|\dot{M}_{\rm eHe}|$) occurring in this age range. 

It is obvious that, in addition to the four cases illustrated in Fig.~\ref{fig_resolu}, all sorts of intermediate situations may be created, depending on how well the mass interval around \Mhef\ is sampled by the stellar evolutionary tracks that are used to build isochrones. It may also be that different sets of evolutionary tracks, when produced at a $\Delta\Mi\simeq0.01$~\Msun\ resolution as our own, will reveal a more gradual change in the evolutionary features across \Mhef. This does not imply that such tracks will avoid the AGB-boosting phase, since as demonstrated in the right panel of Fig.~\ref{fig_resolu}, the AGB-boosting period can be generated even if the CHeB burning lifetimes change over a mass interval as wide as $\Delta\Mi=0.08$~\Msun. More generally, we can affirm that {\em as long as the CHeB burning lifetime increases with mass above \Mhef, in a way that largely compensates for the decrease in main-sequence and RGB lifetimes over the same interval, the AGB-boosting period has to occur}. Fig.~\ref{fig_resolu}, and the simpler arguments in \citet{GirardiBertelli98}, show this clearly. This effect can be eliminated only if we get rid of the increase in CHeB burning lifetimes occurring above \Mhef, or, alternatively, if incomplete descriptions of the light evolution of SSPs (as the one provided by the FCT) are adopted, as shown in our previous Fig.~\ref{fig_fuelt}.


\subsection{Other related results}

From the previous discussion, it is clear that the present results are a consequence of the abrupt changes in the evolutionary features at $\Mi>\Mhef$, as the electron degeneracy no longer develops in the core before He ignition: over a small interval of \Mi, the shortening of the RGB lifetime is followed by an abrupt reduction in the luminosity at which the He-burning phase takes place, that in turns causes a sharp increase of its lifetime. These changes in the CHeB lifetime and luminosity have been found in many other sets of evolutionary tracks since \citet{Sweigart_etal90}, as in e.g.\ \citet{Pols_etal98}, \citet{Dominguez_etal99}, \citet{Castellani_etal00}, \citet{Girardi_etal00}, \citet{Pietrinferni_etal04} and \citet{WeissFerguson09}. The Appendix provides a more extensive comparison between different sets of tracks, revealing that the large increase on the CHeB burning lifetimes above \Mhef\ is not only a common feature in the tracks where the He-burning evolution is computed, but is also expected from the conditions at the stage of He-ignition, in nearly all sets of tracks computed up to that stage. Indeed, it seems well accepted \citep{Castellani_etal00, Dominguez_etal99} that the longer CHeB burning lifetimes slightly above \Mhef\ derive from the smaller core masses (and hence smaller initial He-burning luminosities) at He-ignition, in these stars.  

In addition, we recall that the same changes in core masses and He-burning luminosities at \Mhef\ are at the origin of the appearance of secondary and dual red clumps, that are nowadays well observed in the Milky Way and other Local Group galaxies \citep[][and references therein]{Girardi_etal98, Girardi99, Dalcanton_etal12, Tatton_etal13, Stello_etal13}, and even in a few populous MC star clusters \citep{Girardi_etal09}. 

Therefore, the AGB-boosting effect is not a feature that pertains to a single, isolated set of evolutionary tracks. Moreover, our estimates of the emitted light (or fuel) \textit{as a function of the stellar initial mass} are well in line with the behaviour suggested by \citet{RenziniBuzzoni86}, \citet{Sweigart_etal90} and \citet{Maraston05}.  In fact, the AGB-boosting feature does not take place when we consider the behaviour of the evolutionary tracks as a function of the initial stellar mass (Fig.~\ref{fig_fuelm}). It only shows up if we look at the emitted light \textit{as a function of the age}, provided we can count on stellar isochrones with a high age (and mass) resolution (as in the left panel of Fig.~\ref{fig_fuelt}, and in the bottom panels of Fig.~\ref{fig_resolu}).
 
\section{Consequences in the context of the TP-AGB controversy}
\label{controversy}

The previous analysis has demonstrated the expected appearance of triple TP-AGB branches at ages $\sim\!1.6$~Gyr. We designate this effect with the generic name of ``AGB boosting'' instead of simply ``AGB triplication'', because details about its duration and ``boosting factor'' may well change as more detailed evolutionary tracks are computed. But we emphasize that -- as demonstrated in the previous section -- as long as the CHeB lifetimes sharply increase above \Mhef, an AGB-boosting period {\em has} to occur. 

The AGB-boosting phase is a temporary feature, preceded and followed by a smoothly-varying light contribution from TP-AGB stars. It cannot last much longer than 0.1~Gyr (again depending on details of evolutionary tracks), as we can infer from Figs.~\ref{fig_taum} and \ref{fig_resolu}. In the wider context of galactic evolution, this is a short-lived feature, and one may wonder whether it could be affecting the conclusions of EPS models in any significant way. Should not this feature be completely smeared out when convolved with the continuous star-formation histories typical of star-forming galaxies? 

And indeed, we believe this feature {\em is} smeared out in most nearby galaxies. Its critical importance resides in the fact that it affects, in a significant way, the MC clusters which are {\em the classical calibrators} of the contribution of TP-AGB stars to EPS models.  

\begin{figure}
\includegraphics[width=\columnwidth]{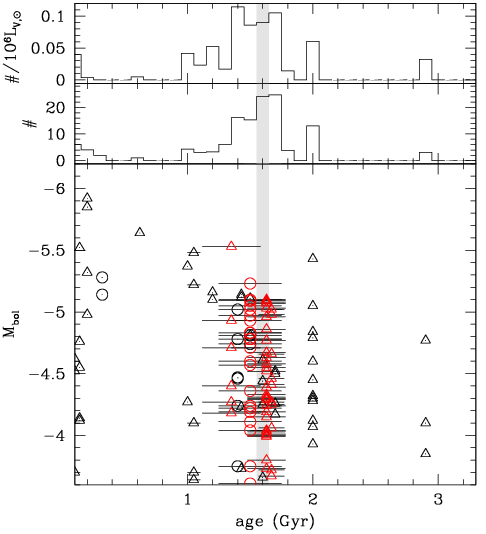}
\caption{The {\bf bottom panel} shows the location of TP-AGB stars in MC clusters in the $\Mbol$ vs.\ age plane, considering just the stars above the tip of the RGB at $\Mbol=-3.6$. Circles are used for stars in SMC clusters, and triangles for the LMC. Stars marked in red belong to the clusters NGC~419, 1806, 1846, 1751 and 1783, for which the error bars indicate the approximate age range of their multiple populations. The {\bf middle panel} shows the age histogram of these TP-AGB stars, whereas the {\bf upper panel} shows the same but for the number of AGB stars normalized to the clusters $V$-band integrated luminosities. In all panels, the light grey area indicates the narrow age interval for which the TP-AGB population is expected to be boosted (cf.~\ref{models}).}
\label{fig_clusters}
\end{figure}

Let us consider Fig.~\ref{fig_clusters}, which shows the TP-AGB stars in MC clusters from the classical compilation by \citet{Frogel_etal90}. The plot shows all the stars above the tip of the RGB at $\Mbol=-3.6$, for which we could attribute ages in the range from 0.1 to 3.3~Gyr. This exclude clusters which, despite containing a few candidate TP-AGB stars, are too young \citep[like NGC~1850, 1854 with $t\!<\!0.1$~Gyr, cf.][]{Pietrzynski_etal98} or too old \citep[like NGC~121, 339, 361, 416, 1841 and Kron~3; cf.][]{Olszewski_etal96, Rich_etal00} to be of interest here. Cluster ages were in general taken from the recent compilation from \citet{Noel_etal13} -- which mostly includes age estimates derived from data that reaches the cluster turn-offs -- but for the following cases: 
\begin{itemize}
\item For the star clusters NGC~411, 419, 1806, 1846, 1751 and 1783, which present clear signs of multiple populations, we adopt the mean ages and their dispersions found in the more recent and detailed CMD analyses based on HST data. More specifically, for NGC~419 we adopt the $1.50\pm0.25$~Gyr age range that encompasses the bulk of its star formation \citep{Rubele_etal10}. The same range is attributed to NGC~411, given the great similarity of its color--magnitude diagram with the NGC~419 one \citep{Girardi_etal13}. For NGC~1751, 1783 and 1846 we adopt the full age ranges found in their centres by \citet{Rubele_etal10, Rubele_etal11}. For the latter clusters, similar age ranges are also obtained by \citet{Goudfrooij_etal09, Goudfrooij_etal11a}, using essentially the same data but different methods and various sets of stellar models. The ages for NGC~1806 are also taken from \citet{Goudfrooij_etal11a}. Notice that for the all these clusters, the mean ages are similar (to within 0.15~Gyr) to those found in the \citet{Noel_etal13} compilation, but for NGC~419 -- the cluster with the most TP-AGB stars -- to which they assign a single age of 1.2~Gyr. The latter is actually the age of the youngest stellar population in NGC~419 as determined by \citet{Glatt_etal09}, and not its mean age.
\item For NGC~152, we take the age from \citet{Rich_etal00}, derived from HST WFPC2 photometry.
\item Finally, for a few other clusters, like NGC~299, 306, 1652, no reliable ages were found. They contain just 9 candidate TP-AGB stars, which certainly do not affect our discussion. 
\end{itemize}
This sample is expected to be more or less unbiased in terms of its age distribution, in the sense that the original selection of clusters by \citet{Frogel_etal90} was not aiming to sample any particular age range, nor any particular range of age-sensitive properties (like the integrated colors).

The bottom panel of Fig.~\ref{fig_clusters} shows the location of the TP-AGB stars in these MC clusters in the $\Mbol$ vs.\ age plane, while the two upper panels show the histogram of stellar ages, using either straight star counts or counts normalized to the cluster $V$-band integrated luminosity -- which represents a rough measure of present cluster masses. All these plots make evident a strong concentration of TP-AGB stars close to the 1.5-Gyr age range. Three factors are at play here:
\begin{enumerate}
\item This age range happens to contain some of the biggest intermediate-age LMC clusters. Classical examples are NGC~419, 1806, 1846, 1751 and 1783, with present masses as high as $1.8\times10^5$~\Msun\ \citep{Goudfrooij_etal11a}, and which contain 69 of the 129 TP-AGB stars plotted in Fig.~\ref{fig_clusters}. Curiously enough, these are clusters that present multiple turn-offs (probably reflecting their large total masses) and dual red clumps \citep{Girardi_etal09}. These observed properties are clearly indicating that their turn-off masses are effectively close to \Mhef, and hence that their assigned mean ages should be reasonably accurate. \label{item_clust}
\item The TP-AGB numbers may indeed be boosted in these $\sim\!1.5$~Gyr clusters, due to the effect discussed in this paper. All the big clusters with multiple turn-offs, indeed, present a substantial superposition between their age ranges and the 1.57--1.66~Gyr boosting period. \label{item_boost}
\item The TP-AGB lifetime is likely to peak at masses of $\Mi\sim2$~\Msun, which broadly corresponds to turn-off ages between $1$ and $2$~Gyr. Several different sets of evolutionary models more or less independently indicate this \citep[e.g][]{MarigoGirardi07, WeissFerguson09}. \label{item_lifetime} 
\end{enumerate}
 
It is the coincidence of these three factors to be particularly insidious. Were MC clusters and their total masses uniformly distributed in age (contrary to item \ref{item_clust}), grouping several clusters in wide age bins would be enough to smear our any short-living evolutionary effect (as the one mentioned in item \ref{item_boost}); but this is apparently not feasible, because the biggest MC clusters are preferentially found in the ``worst possible'' age interval, where the AGB-boosting can fully have its impact. To complicate things, the TP-AGB lifetimes are far from negligible in this mass/age range (cf.~item \ref{item_lifetime}).
 
Given the concentration of big MC star clusters in the AGB boosting period, estimates of the TP-AGB contribution to intermediate-age stellar populations -- either involving numbers/lifetimes as in \citet{GirardiMarigo07}, integrated luminosities as in \citet{Frogel_etal90}, or integrated colours as in \citet{Maraston05} and \citet{Noel_etal13} -- may be biased to too large values. The problem is that these very populous intermediate-age MC clusters are presently not recognized as having a boosted AGB population. Their TP-AGB numbers and integrated luminosities are taken as representative of age intervals much wider than the 0.1-Gyr interval in which the boosting occurs. For instance, in \citet{Noel_etal13} the clusters likely affected by the boosting period are spread into two age bins spanning the complete age interval from 0.9 to 2 Gyr (which, indeed, happen to present the reddest mean integrated colours). The problem is even worse in previous works, in which the cluster age determinations of intermediate-age MC clusters were typically much coarser, so that even wider age intervals could be affected.
 
Therefore, the AGB boosting period very likely causes an overestimation of the TP-AGB contribution, which later propagates into EPS models of galaxies.
In all this matter, the exact age distribution of the stars in the MC clusters NGC~419, 1751, 1806, 1846 and 1783 is really critical. It would be enough to shift the age determination of these clusters by just 10 percent, to greatly increase/decrease the level of coincidence between their TP-AGB ages, and the AGB-boosting period. On the other hand, removing these clusters from the calibration sample used to constrain TP-AGB models would be detrimental, since this would dramatically reduce the numbers of observed TP-AGB stars at our disposal. Moreover, the remaining clusters in the \citet{Frogel_etal90} catalog \citep[see also][]{vanLoon_etal05} typically contain very few TP-AGB stars per cluster (less than 5, except for NGC~1978 which has 12, at an age of 2~Gyr), and hence are strongly affected by stochastic fluctuations in their star counts and integrated colors.

\section{Concluding remarks}
\label{conclu}

The broad implication of our findings is that present models calibrated on MC clusters may be significantly overestimating the TP-AGB flux contribution to models of distant galaxies. Precise numerical estimates of the excess factors are beyond the scope of this paper. Anyway, the way for improving this situation is quite clear to us:

(1) The data for TP-AGB stars in intermediate-age star clusters in the Magellanic Clouds should be carefully revised, in view of re-deriving the lifetimes and integrated flux of their TP-AGB stars. Simply making plots of AGB star counts normalized to the $V$-band cluster luminosity \citep{GirardiMarigo07} and integrated colours \citep{Noel_etal13} versus a rough estimate of cluster age, with clusters binned into wide age bins, is not enough. It is necessary to locate the exact position of each cluster with respect to the AGB-boosting period. Approximations based on the fuel-consumption theorem are also to be avoided. Describing the Post-MS evolution of a given age with a single track with initial mass $M_{\rm TO}$, the FCT is not able, by construction, to cross any evolutionary phase more than once -- nor, more generally, to take into account any temporary increase in the rate of production of stars in post-MS stages due to changes in the post-MS lifetimes \citep[as in the effect described by][]{GirardiBertelli98}. 
In order to correctly take into account the AGB-boosting effect, it is absolutely necessary (a) to model every cluster CMD using detailed isochrones calculated from tracks with a very fine mass resolution, (b) to precisely identify the actual turn-off mass, and its distribution in the clusters with multiple turn-offs. This may not be an easy task. HST-quality photometry is needed, as well as a careful consideration of effects such as incompleteness, crowding, mass segregation, and the field contamination. However, \citet{Rubele_etal10, Rubele_etal11, Rubele_etal13} demonstrate that such an accurate work of model fitting is not beyond reach. 
 
(2) In the absence of a representative sample of star clusters well distributed in age and containing large numbers of TP-AGB stars, the use of galaxy fields to estimate the TP-AGB contribution appears legitimate -- at least, their more continuous star formation history (SFH) would ensure that they are less affected by the AGB-boosting effect than the present samples of MC clusters. The basic requirement for a quantitative work is having reliable estimates of the galaxy SFHs. This is the case of post-starburst galaxies \citep{Kriek_etal10, Zibetti_etal13}, and of many nearby galaxies presently sampled with deep ground-based and HST photometry \citep[see e.g.][]{Gullieuszik_etal08, Girardi_etal10, Weisz_etal11, Melbourne_etal12}. However, it is also clear that when looking at entire galaxies, one will inevitably miss many of the details of the TP-AGB evolution that we would like to constrain.

Needless to say, these are all steps we are going to pursue in forthcoming papers.

\acknowledgments
We warmly thank Julianne Dalcanton for her great encouragement to this work, Stephane Charlot for the useful comments, and Stefano Rubele and Leandro Kerber for showing us that a better approach to studying Magellanic Cloud clusters is feasible. Many thanks also to the users of our isochrones who pointed out to their ``weird features'' with genuine curiosity rather than with distrust. We acknowledge financial support from contract ASI-INAF n. I/009/10/0, from Progetto di Ateneo 2012, University of Padova, ID: CPDA125588/12, and from PRIN 2009.

\appendix

\section{A. Are our results reliable?}

At the request of the referee, we provide here additional details about our models, and comparisons with other authors. They are mostly aimed at testing the reliability of our results, especially in regard to the sharpness of the transition in evolutionary features taking place at $\Mi=\Mhef$. Before proceeding, we should emphasize that our models provide {\em by far} the best mass resolution available in the literature. Any comparison with other authors is very much limited by this fact.

\begin{figure}
\includegraphics[width=\columnwidth]{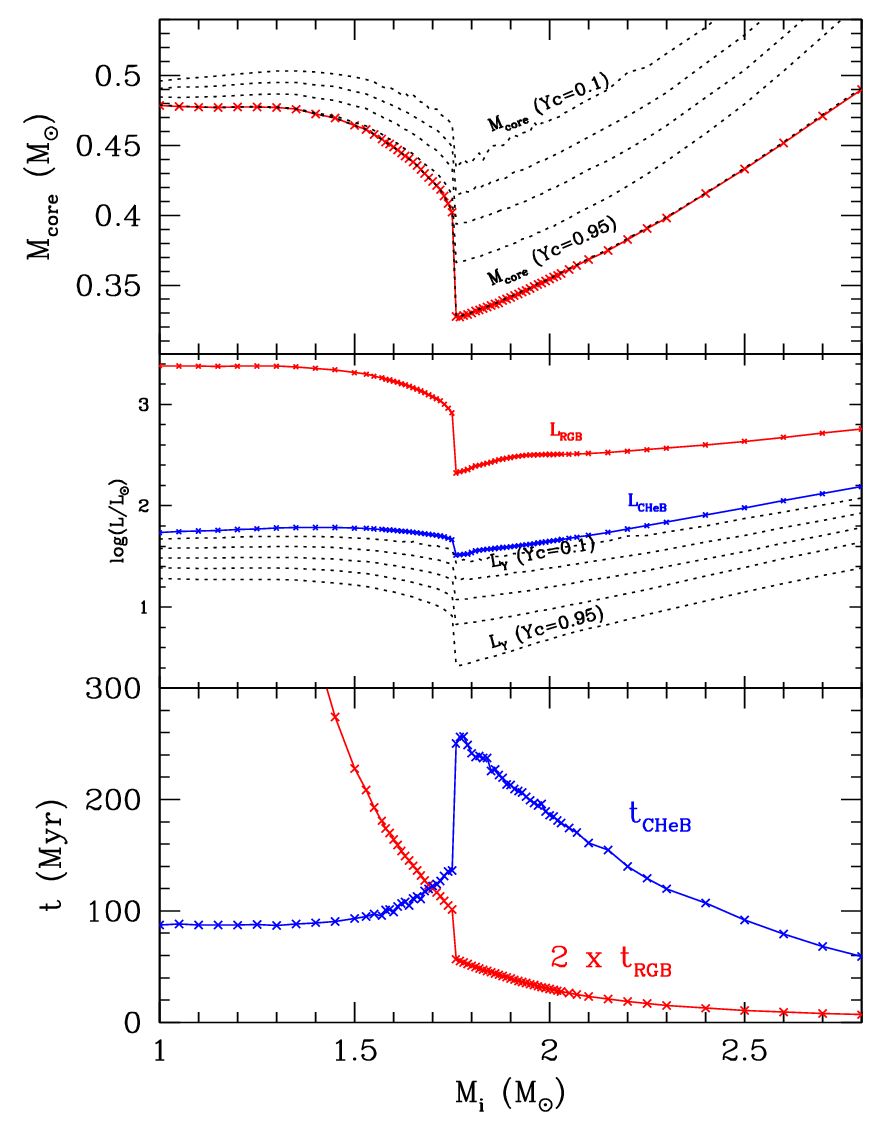}
\caption{Several properties derived from our fine grid of $Z=0.006$ evolutionary tracks, as a function of initial mass. The crosses represent the values for tracks actually computed, which are then connected by straight lines. {\bf Bottom panel:} The total lifetimes in the RGB (red) and CHeB (blue) phases as a function of mass.  {\bf Middle panel:} The maximum luminosities reached at the stage of He-ignition, denoted as $L_{\rm RGB}$ (red), and minimum luminosities reached at the stage of quiescent He-burning, denoted as $L_{\rm CHeB}$ (blue). In addition, the luminosities provided by the He-burning core at different stages of the CHeB evolution are plotted with dotted lines (for $Y_{\rm c}=0.95$, 0.7, 0.5, 0.3, and 0.1, from bottom to top). {\bf Top panel:} The mass of the H-exhausted core at the beginning of the CHeB phase (red). The dotted lines denote the increased core mass at successive stages of the CHeB evolution (for $Y_{\rm c}$ decreasing from 0.95 to 0.1 from bottom to top, as before).}
\label{fig_ourlifetimes}
\end{figure}

\subsection{A.1 Properties of our models}

As already discussed, the AGB-boosting owns its origin to the sharp increase in CHeB burning lifetimes at masses close to \Mhef. This increase is further illustrated in the bottom panel of Fig.~\ref{fig_ourlifetimes}, which details how the relevant lifetimes change with the stellar initial mass. More specifically, we plot the run of the ``RGB lifetime'', $t_{\rm RGB}$, measuread as the time span between the H-exhaustion in the core  ($X_{\rm c}=0$) and the maximum luminosity reached soon after He-ignition, and the run of the He-burning lifetime, $t_{\rm CHeB}$, measured as the time span between the He-ignition and the He-exhaustion in the core ($Y_{\rm c}=0$).  It can be seen that both quantities gradually change as \Mi\ approaches the \Mhef\ limit, but the change becomes rather abrupt as \Mhef\ is crossed. Similar changes occur also in the maximum and minimum luminosities reached at the initial stages of He-burning, denoted as $L_{\rm RGB}$ and $L_{\rm CHeB}$. 

It is remarkable that the quantities at the RGB tip, $t_{\rm RGB}$ and $L_{\rm RGB}$ (both marked in red), are derived from stellar models computed on a consistent way from the pre-main sequence up to the He-ignition, therefore they are {\em not affected} by the artificial method performed to build quiescent He-burning models of masses $\Mi\!<\!\Mhef$. The same applies to the core mass at He-ignition, which is measured at a stage in which it is still not altered by the He-burning. Nonetheless, all models indicate the same discontinuity at $\Mi\!=\!\Mhef$, namely a sharp decrease in both $t_{\rm RGB}$, $L_{\rm RGB}$, and $M_{\rm core}$. This sharp decrease is clearly indicating a discontinuity caused by the stellar physics, rather than by the numerical algorithms adopted to skip the He-flash in low-mass models.

\begin{figure}
\includegraphics[width=\columnwidth]{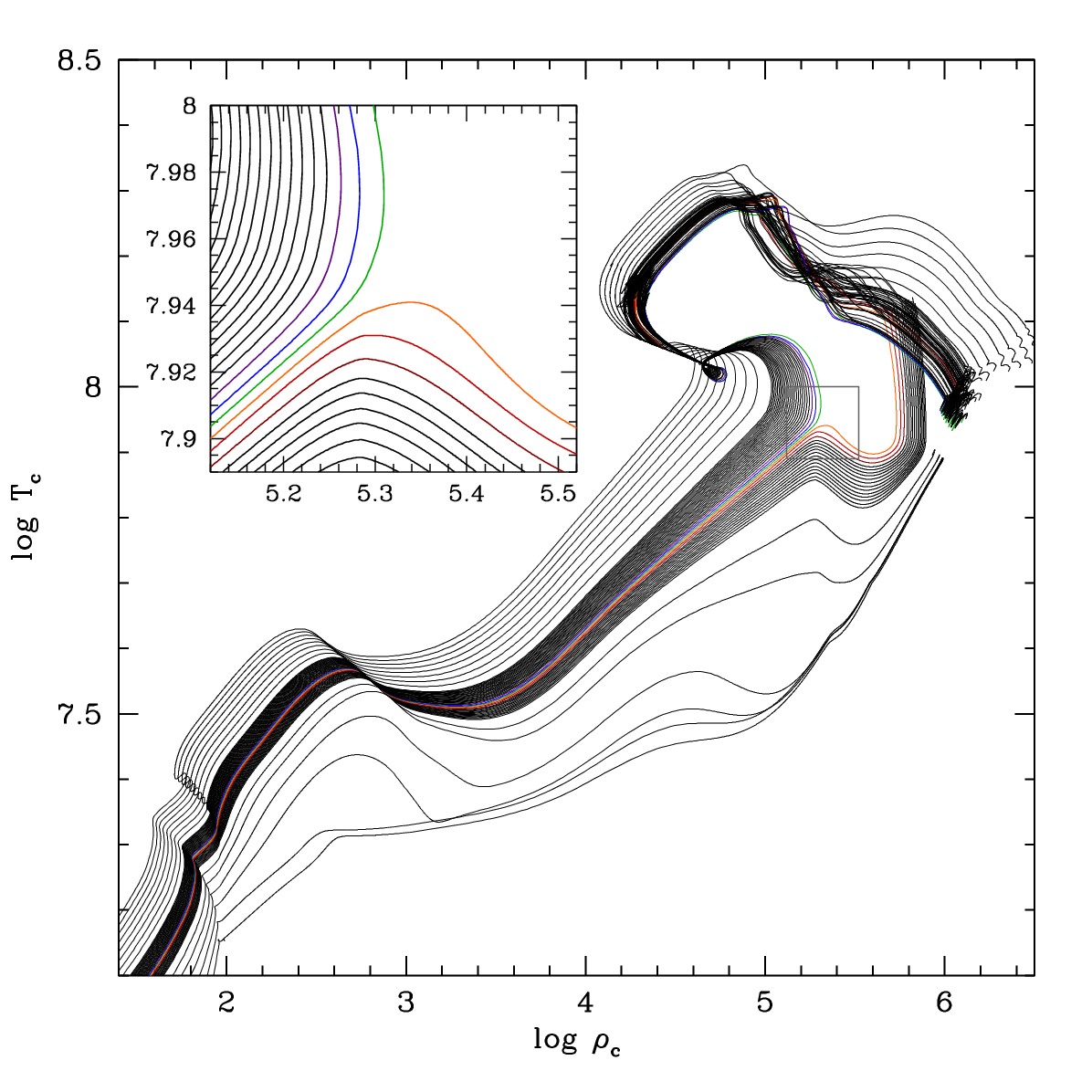}
\caption{The evolution of central temperature vs.\ central density for a subset of our tracks. The central bundle of models are those with initial masses between 1.8 and 2.0~\Msun, separated by steps of 0.01~\Msun\ (from bottom to top).  The more sparse models to both sides of this central bundle cover the mass interval between 1.0 and 2.8~\Msun, at steps of 0.1~\Msun. The central models, colored from maroon to purple, are those with masses between 1.73 and 1.78~\Msun, respectively. The inset details the point at which these models split, before igniting helium, going either towards quiescent He-ignition (to the top-left, for $\Mi\!\geq\!1.76$~\Msun) or to settling electron-degeneracy in their cores (to the bottom-righ, for $\Mi\!\leq\!1.75$~\Msun).}
\label{fig_bifu}
\end{figure}

And indeed, the diagram of central temperature vs.\ central density shown in Fig.~\ref{fig_bifu} shows a clear bifurcation in the evolutionary paths, that sharply separates the models with $M\geq1.76$~\Msun\ from those with $M\leq1.75$~\Msun. The former ignite helium in non-degenerate conditions, while the latter decisely cool at increasing density, delaying He-ignition to a stage further along in the RGB. This bifurcation has been found in numerous sets of published stellar tracks in the past. What is unusual is just to show it with a mass resolution as small as $\Delta\Mi=0.01$~\Msun, as in our case.

After He-ignition, low-mass models have their evolution interrupted and continued from a zero-age-horizontal branch model built with a suitable mass and chemical composition (see Sect.~\ref{models}), while intermediate-mass models are continuously computed until the TP-AGB phase. This dichotomy in the way models are evolved could create additional differences between the two mass ranges (in addition to those caused by the onset of electron degeneracy), because the convective regions developed in the core during He-ignition could have been treated differently in the two cases.  
Especially worrying would be if the overshooting scheme adopted in the tracks were to extend the convective cores in the models just massive enough to survive the He-flash. However, inspection of the models shows that the convective cores of low-mass CHeB models are larger than those of He-burning stars above the $\Mi\!=\!\Mhef$ discontinuity (top panel of Fig.~\ref{fig_ourlifetimes}). This is simply due to the larger H-exhausted cores of the former models, which are a common feature, irrespective of the adopted efficiency of core overshoot and of the way they have been computed. In spite of that, in models with mass just above the discontinuity the He-burning lifetimes is about twice that of the $\Mi<\Mhef$  models. Again this feature is common to all models -- irrespective of the efficiency of core overshooting -- and due to the lower He-burning luminosity which is needed to sustain a smaller He core, i.e.\ to the He-core mass--luminosity relation. The latter can be appreciated by looking at the luminosity coming from He-burning reactions, $L_Y$, at several stages of the CHeB evolution (middle panel of Fig.~\ref{fig_ourlifetimes}).  $L_Y$ is indeed dramatically smaller for the CHeB models with the core mass close to the $0.33$~\Msun\ minimum, soon after He-ignition (as for the curves with $Y_{\rm c}\!=\!0.95$ and  $Y_{\rm c}\!=\!0.7$). At later stages of CHeB, the core masses grow and the differences between the $L_Y$ values between tracks of varying masses become smaller (as seen in the top and middle panels of Fig.~\ref{fig_ourlifetimes}), but anyway the very different $L_Y$ at the start of the CHeB ensure the much longer lifetimes of the models slightly above $\Mi\!>\!\Mhef$. 

Thus, no matter the way the models have been computed, a discontinuity in the He core masses gives rise to a discontinuity in the He-burning lifetimes, with models possessing smaller cores having significantly prolonged $t_{\rm CHeB}$.


\begin{figure}
\includegraphics[width=\columnwidth]{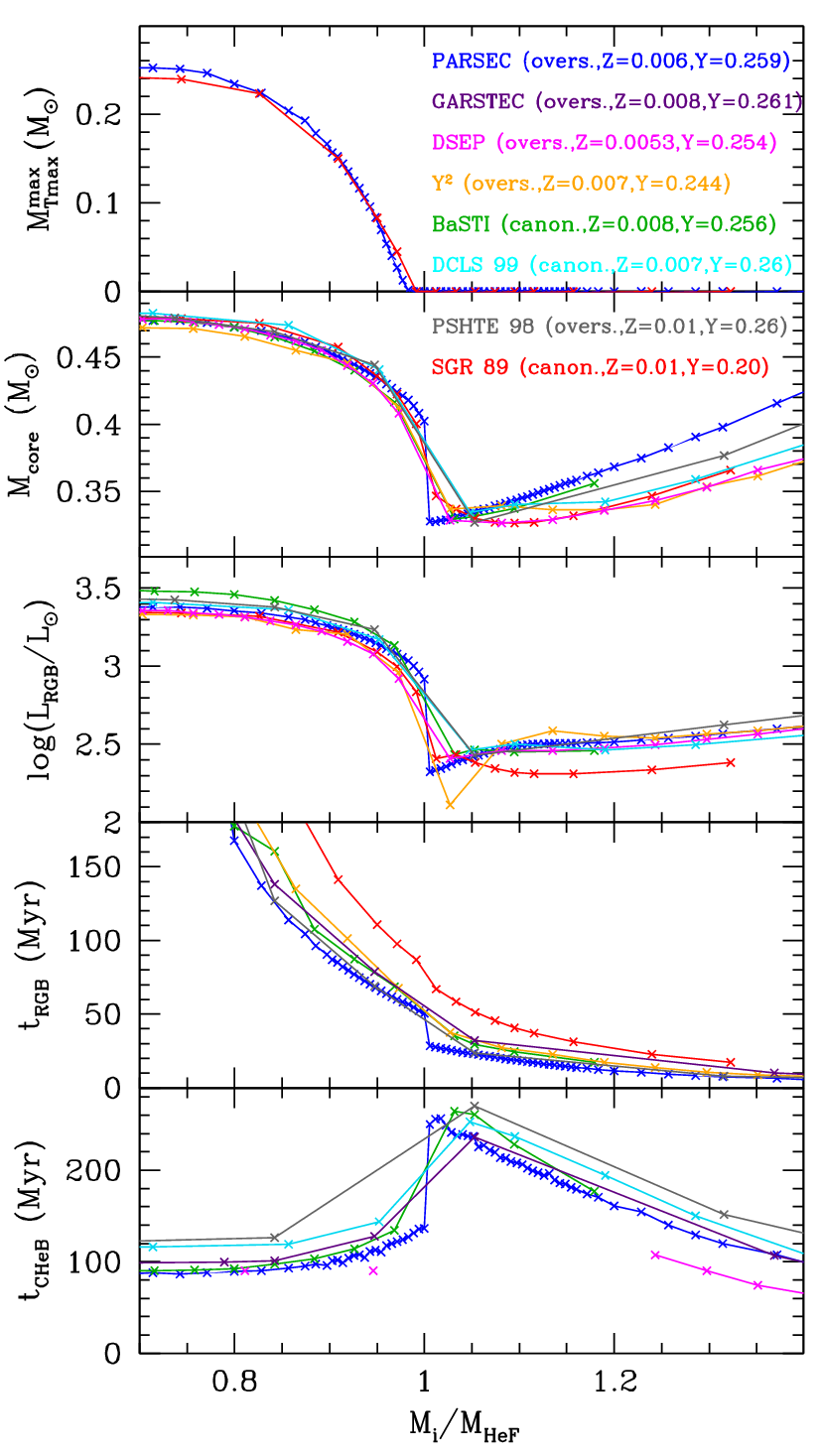}
\caption{Results from several sets of stellar models, comparing quantities that are directly related to the changes in the stellar lifetimes occurring in the vicinity of \Mhef. Since different models are computed with different efficiencies of overshooting, the abscissa shows the mass divided by the \Mhef\ value appropriatte for each set. The crosses are models actually computed by the several authors. Panels from top to bottom present: the maximum mass cordinate of the point of maximum temperature inside the star, before He ignition, $M_{T{\rm max}}^{\rm max}$; the core mass at He-ignition, $M_{\rm core}$; the maximum luminosity soon after He-ignition, $L_{\rm TRGB}$; the total RGB plus subgiant branch lifetime, $t_{\rm RGB}$; and the total core He-burning lifetime, $t_{\rm HeB}$. The models are taken from 
\citet[][SGR~89]{Sweigart_etal89}
\citet[][PSHTE~98]{Pols_etal98}
\citet[][DCLS~99]{Dominguez_etal99}
\citet[][BaSTI]{Pietrinferni_etal04}
\citet[][Y$^2$]{Demarque_etal04}
\citet[][DSEP]{Dotter_etal08}
\citet[][GARSTEC]{WeissFerguson09},
and from this work (PARSEC).}
\label{fig_otherlifetimes}
\end{figure}

\subsection{A.2 Comparison with other stellar models}

The abrupt changes occurring in the evolutionary features and lifetimes in the vicinity of \Mhef\ have been already discussed by many authors, using grids of stellar models computed with different input physics and various degrees of completion and mass resolution. Some of these models, chosen among the most complete available in the literature, and with initial chemical composition similar to the LMC one, are compared to ours in the different panels of Fig.~\ref{fig_otherlifetimes}. The comparison is performed as a function of $\Mi/\Mhef$, because convective overshooting has the effect of systematically increasing the core masses and hence of reducing the value of \Mhef\ \citep[see e.g.][]{Sweigart_etal90, Girardi_etal00}. In other words, we will be using $\Mi/\Mhef$ as a proxy for the core mass developed after the main sequence, so that we can directly compare -- at least in the vicinity of \Mhef\ -- models which naturally develop very different core masses. Since the definition of \Mhef\ might slightly differ from author to author, we simply redefine \Mhef\ for each set, as being the mass value that presents the fastest variation of the quantity $L_{\rm RGB}$ with mass.

\citet{Sweigart_etal89, Sweigart_etal90} were the first to explore the issue of how fast is the transition in evolutionary features across \Mhef, in the context of the long-sought ``RGB phase transition'' defined by \citet{RenziniBuzzoni86}. They computed a series of canonical (without overshooting) models at a typical separation of $\Delta\Mi=0.05$~\Msun, but only up to the He-ignition. It can be seen that the behaviour of the present models is quite similar to those of \citet{Sweigart_etal89}, apart from modest offsets in the plotted luminosities and lifetimes, and for \citet{Sweigart_etal89} models suggesting a more gradual variation of the stellar properties for masses above $\Mhef$. The offsets in luminosities and lifetimes are no surprise, given the differences in the initial chemical composition, and the large changes in the physical input adopted by stellar models in the last two decades.

Particularly interesting is the comparison presented at the top panel, which presents the maximum mass coordinate of the maximum temperature reached inside the star before the He-flash, $M_{T{\rm max}}^{\rm max}$. A value larger than zero is usually interpreted as the signature of an electron-degenerate core, with its temperature inversion caused by the efficient conduction and by neutrino cooling. The mass at which $M_{T{\rm max}}^{\rm max}$ becomes null has sometimes been used as a definition for \Mhef\ \citep[e.g.][]{Castellani_etal00}. It is interesting to note that in our models $M_{T{\rm max}}^{\rm max}$ becomes null already at $\Mi=1.72$~\Msun, which is at least 0.03~\Msun\ smaller than the mass value for which the evolutionary features more rapidly change, \Mhef. Such a distinction between limiting masses could not have been detected in \citet{Sweigart_etal89} models, given their $\Delta\Mi=0.05$~\Msun\ resolution. Apart from this fine detail, the run of $M_{T{\rm max}}^{\rm max}$ with mass is surprisingly similar between these two sets of models.

The behaviour of \citet{Sweigart_etal89} models seem to be confirmed by those from \citet{Pols_etal98} and \citet{Dominguez_etal99}, although the latter are computed with a slightly worse resolution in mass. On the other hand, \citet{Pols_etal98} and \citet{Dominguez_etal99} compute the complete He-burning evolution for models below and above the $\Mhef$ limit, finding that the CHeB lifetime more than doubles for stars slightly more massive than $\Mhef$ -- as also found in our models.  Very similar behaviour is also presented by the GARSTEC tracks from \citet{WeissFerguson09}.

Also suggestive is the comparison between our models and some recent ones such as the Yale-Yonsei \citep{Demarque_etal04}, BaSTI \citep{Pietrinferni_etal04}, and DSEP \citep{Dotter_etal08}. These sets of models predict a quite similar evolution of the quantities at the He-ignition, although, in all these cases, the finest details of the transition at \Mhef\ are missed because of the limited mass resolution (of about $\Delta\Mi=0.05$~\Msun\ in the best case).

In the case of BaSTI models, the comparison can be extended to the He-burning luminosities and lifetimes, which compare quite well. In BaSTI, the longest CHeB burning lifetimes are found just a few hundredths of solar masses above \Mhef\ -- although this trend is not very clear give the scarcity of tracks computed in the immediate vicinity of \Mhef. Another interesting point is that the BaSTI tracks present the same trend of presenting the smallest luminosities at He-ignition in the mass interval immediately above \Mhef. This behaviour is probably reflecting the minimum in the core mass at He-ignition, which is also found in the same interval. Similar trends are also observed in the DSEP models, which behave in a way very similar to BaSTI ones. Unfortunately, DSEP does not contain He-burning models in the immediate viciniy of \Mhef, and largely fail to sample the mass interval for which He-burning lifetimes are expected to exceed 100~Myr.

Overall, we find a good level of agreement between the behaviour of our models and others found in the literature, despite the great differences in parameters such as the overshooting efficiency, initial metallicity and helium content. Since our models have by far the best mass resolution among these different sets, and are the most complete in the calculation of He-burning phases, it is natural that they provide the most detailed description of the $\Mi\simeq\Mhef$ interval. As surprising as it can be, they also provide the sharpest transition in the stellar evolutionary features across this mass interval.

$\frac{}{}$


\end{document}